\documentclass[aps,preprint,superscriptaddress,floatfix,nofootinbib]{revtex4-2}

\usepackage{amsmath,amssymb}
\usepackage{graphicx}
\usepackage{bm}
\usepackage{xcolor}
\usepackage[caption=false]{subfig}
\usepackage{hyperref}
\hypersetup{colorlinks=true,linkcolor=blue!60!black,citecolor=blue!60!black,urlcolor=blue!60!black}

\usepackage{xcolor}
\newif\ifrev
\revfalse

\ifrev
  \newcommand{\rev}[1]{\textcolor{blue}{#1}}
  \newcommand{\revc}{\color{blue}}
\else
  \newcommand{\rev}[1]{#1}
  \newcommand{\revc}{}
\fi

\begin{document}

\title{Learning Cole-Hopf Coordinates for Stochastic Growth Dynamics: the KPZ class}

\author{Lorenzo Mensi}
\affiliation{Universit\'e Paris-Saclay, CNRS, LISN, 91400 Orsay, France}
\author{Simone Cuonzo}
\affiliation{Dipartimento di Fisica, Sapienza Universit\`a di Roma, Piazzale Aldo Moro 5, 00185 Roma, Italy}
\author{Maria Chiara Angelini}
\affiliation{Dipartimento di Fisica, Sapienza Universit\`a di Roma, Piazzale Aldo Moro 5, 00185 Roma, Italy}
\author{Sergio Chibbaro}
\affiliation{Universit\'e Paris-Saclay, CNRS, LISN, 91400 Orsay, France}
\author{Cyril Furtlehner}
\affiliation{Inria-Saclay,Universit\'e Paris-Saclay, LISN, 91400 Orsay, France}
\author{Alberto Rosso}
\affiliation{Universit\'e Paris-Saclay, CNRS, LPTMS, 91400 Orsay, France}

\begin{abstract}

Inferring effective equations from observations of nonlinear stochastic
processes is challenging because individual trajectories are noisy
realizations, while the relevant physical information is encoded in ensemble
statistics. We introduce a data-driven framework that learns a change of
variables in which nonlinear stochastic dynamics become linear. The model
combines a neural change of representation with a linear stochastic latent
evolution and is trained through a short-time moment-matching objective,
thereby avoiding trajectory \rev{simulation} during optimization. This construction
yields an interpretable generative model: both the learned transformation and
the latent evolution operators can be directly inspected.

We validate the method on the one-dimensional Kardar--Parisi--Zhang (KPZ)
equation, a paradigmatic model of stochastic growth for which the
Cole--Hopf transformation maps the nonlinear dynamics to the stochastic heat
equation. From simulated height trajectories alone, the model recovers the
local exponential Cole--Hopf map and the associated linear stochastic operator.
Generated trajectories reproduce the main statistical signatures of the KPZ
universality class, including the growth law $\rev{\mathrm{Var}(h^{\rev{t}})}\sim t^{2\beta}$ with
$\beta=1/3$ and the GOE Tracy--Widom height distribution for flat initial
conditions. These results show that interpretable stochastic equations can be
inferred from data by learning representations in which the dynamics closes in a simple form. Furthermore the development of the physical informed learning permits to get insights on the physics of the dynamical process.

\end{abstract}

\maketitle

\section{Introduction}

Many dynamical processes in nature are intrinsically stochastic, from
microscopic thermal fluctuations to macroscopic growth phenomena. In such
systems, the observed trajectories are noisy realizations of an underlying
probabilistic dynamics, and the relevant physical information is encoded not
in a single path but in statistical observables such as correlations and
probability distributions. Inferring an
effective stochastic equation from data therefore remains a central problem
across statistical physics and related fields~\cite{gardiner2004stochastic,
kubo2012statistical,marconi2008fluctuation}.

Traditionally, this task relies on physical insight. One proposes an effective
Langevin equation or stochastic partial differential equation, guided by
symmetries, conservation laws, locality, and dimensional arguments, and then
validates the model \emph{a posteriori} by comparing its predictions with
experiments or simulations. A paradigmatic example is the
Kardar--Parisi--Zhang (KPZ) equation~\cite{KPZ1986,kardar2007statistical},
which describes the stochastic evolution of the height $h(x,t)$ of a growing
one-dimensional interface,
\begin{equation}
\partial_t h(x,t)
=
\nu\,\partial_x^2 h(x,t)
+
\frac{\lambda}{2}\left(\partial_x h(x,t)\right)^2
+
\sqrt{2D}\,\eta(x,t),
\label{eq:kpz_intro}
\end{equation}
where the first term smooths the interface, the second accounts for lateral
growth, and $\eta(x,t)$ is a Gaussian space-time white noise representing
random fluctuations. Despite its deceptively simple form, the KPZ equation
captures the universal large-scale behavior of a broad class of
non-equilibrium growth processes. KPZ scaling and fluctuations have been
observed or proposed in systems ranging from turbulent liquid crystals and
polariton condensates to bacterial colonies and other kinetic roughening
phenomena~\cite{TakeuchiSano2010,TakeuchiSano2011,Fontaine2022polariton,
HalpinHealyTakeuchi2015,Takeuchi2018appetizer,Vicsek1990bacterial}.
While this phenomenological approach has been remarkably successful, it
relies on proposing the appropriate evolution equation from physical intuition.
Determining the correct form of the dynamics is often the most challenging
step, especially for complex systems where the underlying mechanisms are only
partially understood.

This difficulty is amplified by nonlinearity. Most complex phenomena are
described by nonlinear evolution equations~\cite{bohr1998dynamical}, for
which exact analytical treatments are rare. In some special cases, however,
a suitable change of variables can drastically simplify the dynamics. A
classical example is the Cole--Hopf transformation, which maps the viscous
Burgers equation to the linear heat equation~\cite{Hopf1950,Cole1951}. In the
stochastic setting, the same idea applies to the KPZ equation: after the
transformation
\begin{equation}
z(x,t)=\exp\!\left[\frac{\lambda}{2\nu}h(x,t)\right],
\end{equation}
the nonlinear KPZ equation is mapped onto the linear stochastic heat equation
with multiplicative noise~\cite{BertiniGiacomin1997,Hairer2013solving},
\begin{equation}
\partial_t z(x,t)
=
\nu\,\partial_x^2 z(x,t)
+
\frac{\lambda}{2\nu}\sqrt{2D}\,z(x,t)\,\eta(x,t),
\end{equation}
\rev{where the noise is interpreted using the Stratonovich convention.} KPZ therefore provides an example of a nonlinear stochastic field theory
that admits an exact linearizing transformation. This naturally raises the
following data-driven question: can such simplifying transformations be
learned directly from stochastic trajectories?

Machine learning provides a possible route to this problem. Neural
differential-equation models have made it possible to learn continuous-time
dynamics from data, and neural stochastic differential equations extend this
idea to stochastic trajectories~\cite{chen2018neural,TzenRaginsky2019,
li2020scalable,Kidger2021NeuralSDE}. Recent neural operator and neural SPDE
approaches further address stochastic spatiotemporal dynamics at the level of
solution operators~\cite{Salvi2022NeuralSPDE}, although they currently remain
restricted to specific classes of systems. In parallel, sparse-regression methods such as SINDy and PDE-FIND seek to infer governing equations directly from data~\cite{Brunton2016SINDy,Rudy2017PDEFind,Schaeffer2017PDE}. \rev{These were originally formulated for deterministic dynamics, but the same sparse-identification idea has been extended to stochastic systems by combining it with Kramers--Moyal estimates of the drift and diffusion coefficients~\cite{Boninsegna2018}.} Together, these developments
show that increasingly complex dynamical systems can now be learned directly
from observations.

For deterministic dynamics, several methods have been developed to learn
interpretable representations in which the evolution becomes simpler. These
include Koopman-inspired linear embeddings~\cite{Mezic2005,Williams2015,
Lusch2018,OttoRowley2019} and autoencoder-based approaches combined with
sparse regression, such as SINDy-AE~\cite{Champion2019datadriven}. Such
methods can identify latent coordinates in which nonlinear deterministic
systems evolve approximately linearly or admit a sparse description, \rev{i.e. governing equations built from only a handful of terms}. These
results demonstrate that machine learning can go beyond \rev{black-box} prediction, \rev{and it can be used to} recover
compact mathematical representations of complex dynamics.

Extending these ideas to stochastic systems is considerably more challenging.
In a deterministic system, a model can be trained by requiring its predicted
trajectory to remain close to the observed one. In a stochastic system, this
criterion is no longer appropriate: even the correct model would generate a
different trajectory if driven by a different realization of the noise.
Instead, the relevant information is encoded in the statistics of short-time
transitions, such as their conditional mean and covariance, rather than in
individual trajectories.

Several recent works have addressed this challenge from different
perspectives.  \rev{Stochastic Force Inference~\cite{Frishman2020} estimates drift and diffusion fields from short-time increments by projection onto a known basis, yielding directly interpretable equations, and has been extended to spatially extended systems through a dissipation-based dimensional reduction~\cite{Gnesotto2020}.} Heck \emph{et al.}~\cite{Heck2025noise} introduce
moment-matching constraints within a variational latent-SDE framework, where
both drift and diffusion are represented by neural networks. Hasan
\emph{et al.}~\cite{Hasan2021identifying} focus on learning latent
representations within a variational autoencoder, while simulation-free
approaches such as SDE Matching~\cite{Bartosh2025sdematching,Song2021}
avoid trajectory integration during training. Operator-theoretic methods
instead learn the infinitesimal generator directly from observations without
introducing an explicit latent representation
~\cite{Kostic2024generator,Klus2020generator}.

These approaches have demonstrated impressive generative performance.
However, their primary objective is to reproduce the observed stochastic
dynamics rather than to recover an interpretable mathematical description of
it. The latent dynamics is typically represented by highly expressive neural
networks, making it difficult to extract an explicit stochastic equation that
can be inspected and analyzed. Our objective is different: rather than merely
learning to generate trajectories, we seek to learn a change of variables in
which the observed dynamics becomes a linear stochastic differential equation
whose coefficients can be read directly from the trained model.

In this work, we introduce a neural framework designed to recover interpretable
stochastic equations from data. The central idea is to learn a local change of
variables in which the observed dynamics becomes a linear stochastic evolution.
The model combines an autoencoder mapping the observed field to a latent
representation with a linear latent SDE, and is trained by matching short-time
increment statistics rather than by rolling out full trajectories during
optimization. This yields a model that is both generative and interpretable:
it can produce new trajectories with the statistics of the training data, while
providing explicit objects---the learned transformation, the latent linear
operator, and the latent noise structure---from which an effective stochastic
equation can be reconstructed.

We validate the framework on the one-dimensional KPZ equation, for which the
Cole--Hopf transformation maps the nonlinear dynamics to the linear stochastic
heat equation~\cite{BertiniGiacomin1997,Hairer2013solving}. This benchmark
allows us to test simultaneously whether the method recovers the correct
linearizing transformation and whether the learned generative model reproduces
the universal statistical properties of the KPZ class. We show that the model
recovers both the Cole--Hopf map and the associated linear stochastic dynamics
directly from height trajectories, while reproducing highly non-trivial KPZ
signatures, including the $t^{2\beta}$ roughening law and the GOE
Tracy--Widom height distribution for flat initial conditions~\cite{
SasamotoSpohn2010,AmirCorwinQuastel2011,calabrese2010free,
CalabreseLeDoussal2011flat,Corwin2012review,QuastelSpohn2015}.

The remainder of the manuscript is organized as follows. In Sec.~II we describe the numerical generation of the KPZ dataset used throughout the paper. In Sec.~III we introduce the interpretable latent-SDE framework, including the local change of variables, the linear latent dynamics, the moment-matching loss, and the training procedure. We then assess the generative performance of the learned model and analyze the learned representation, showing that it recovers both the Cole--Hopf transformation and the associated linear stochastic operator. We also discuss in Sec. ~IV an extension of the same statistical-learning principle to a more flexible autoencoder architecture. Finally, Sec.~V contains our conclusions and perspectives.
\section{Numerical generation of KPZ trajectories}
\label{sec:numerical_kpz}

\begin{figure}[h]
\centering
\includegraphics[width=0.49\columnwidth]{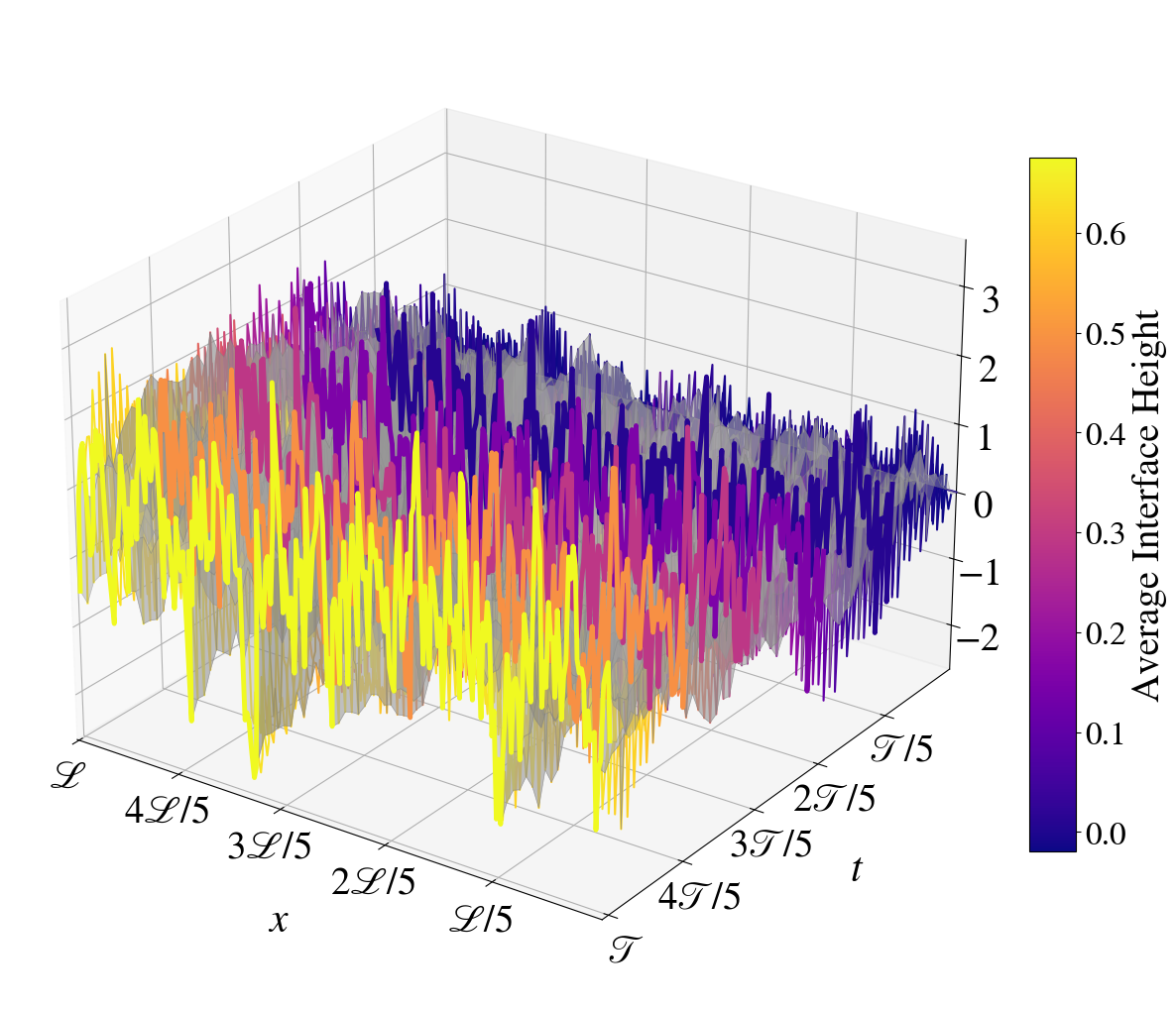}
\includegraphics[width=0.49\columnwidth]{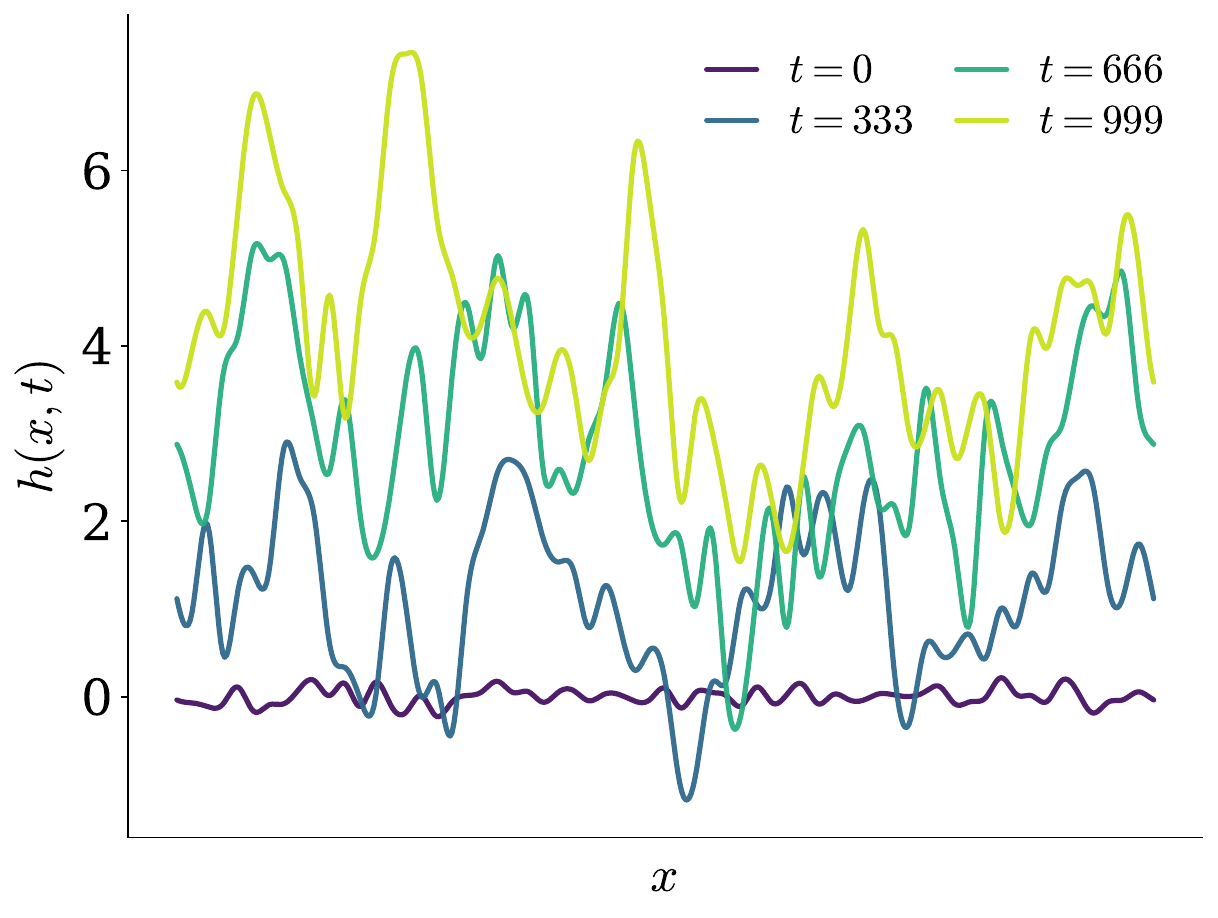}
\caption{
Numerical KPZ trajectories generated from a flat initial condition.
Left: height field $h(x,t)$ as a function of position $x$ and time $t$ for a
single noise realization. Right: corresponding interface profiles at selected
times. The full dataset consists of independent stochastic trajectories
generated from the same flat initial condition and driven by different noise
realizations. The ensemble displays the characteristic roughening and
fluctuation growth of one-dimensional KPZ dynamics.
}
\label{fig:hero-dataset}
\end{figure}

As discussed in the Introduction, the training data are generated from the
Cole--Hopf representation of the one-dimensional KPZ equation. Rather than
discretizing the nonlinear KPZ equation directly, we simulate the corresponding
linear stochastic dynamics for the transformed variable $z$ on a periodic
one-dimensional lattice. \rev{This guarantees that an exactly linear latent evolution underlies the training data, since on the discretized lattice the Cole-Hopf mapping holds only to
leading order in $\Delta x$.} The resulting lattice stochastic equation, \rev{using Itô's formalism}, is

\begin{equation}
dz_i^{\rev{t}} =
\left[
\nu\frac{z_{i+1}^{\rev{t}}+z_{i-1}^{\rev{t}}-2z_i^{\rev{t}}}{\Delta x^2}
+
\frac{D\alpha^2}{\Delta x}z_i^{\rev{t}}
\right]dt
+
\alpha\sqrt{\frac{2D}{\Delta x}}
z_i^{\rev{t}}\sqrt{dt}\xi_i^{\rev{t}},
\label{eq:z_lattice_sde}
\end{equation}

\rev{where $\alpha = \lambda/(2\nu)$ and}  $\xi_i^{\rev{t}}\sim\mathcal N(0,1)$ are independent Gaussian random variables
with zero mean and unit variance. Equivalently, the quantity
$\sqrt{dt} \xi_i^{\rev{t}}$ represents a Brownian increment.

The height field is recovered through the inverse Cole--Hopf transformation

\begin{equation}
h_i^{\rev{t}}=\frac{1}{\alpha}\log z_i^{\rev{t}}.
\label{eq:inverse_ch}
\end{equation}

The deterministic contribution proportional to $z_i$ in
Eq.~\eqref{eq:z_lattice_sde} is the standard It\^o correction associated with
multiplicative noise. Because the noise amplitude itself depends on $z_i$,
fluctuations contribute not only to the variance of the process but also to
its average evolution, generating an additional drift term proportional to
$z_i$.
This correction is necessary for the reconstructed field
Eq.~\eqref{eq:inverse_ch} to evolve according to the KPZ equation. Without it,
the interface would acquire an additional deterministic growth velocity,
while its fluctuation properties would remain unchanged
\cite{BertiniGiacomin1997,Hairer2013solving}. All trajectories are initialized from a flat interface,
\begin{equation}
h_i^{\rev{0}}=0,
\qquad
z_i^{\rev{0}}=1.
\label{eq:flat_ic}
\end{equation}

Unless stated otherwise, we use

\begin{equation}
\nu=0.5,
\qquad
\lambda=1,
\qquad
D=0.5,
\qquad
\Delta x=1,
\qquad
dt=10^{-2},
\qquad
N=256.
\label{eq:simulation_parameters}
\end{equation}
The dataset consists of $10^{\rev{5}}$ independent trajectories generated from
different realizations of the noise (Fig.~\ref{fig:hero-dataset}). \rev{For the parameters used here, the microscopic lattice time
$t_{\rm micro}\sim\Delta x^2/\nu\simeq2$ is not well separated from the
estimated EW--KPZ crossover time $t_c\simeq7.9$ \cite{Krug1997}; consequently, no extended
EW scaling plateau is resolved.}

\section{Physical interpretable model}
\subsection{Architecture}\label{subsec_arch}

\begin{figure}[h] \centering \includegraphics[width=0.9\textwidth]{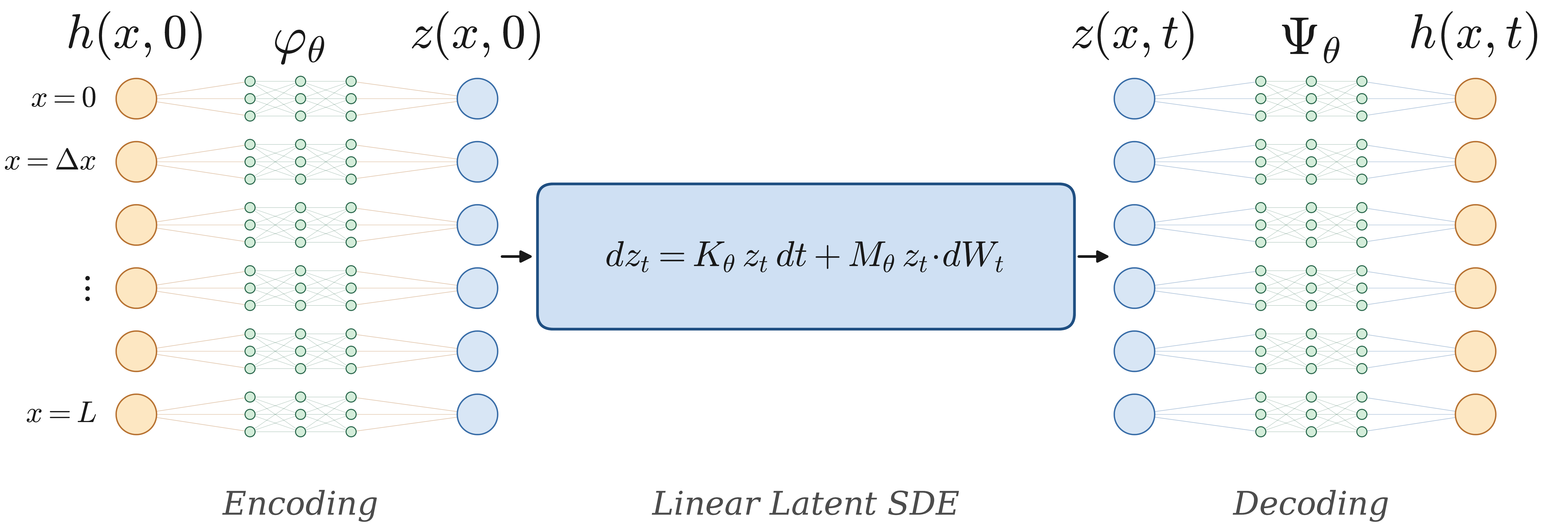} \caption{
Schematic of the model. The observed field $h(x,t)$ is mapped site by site to
a latent variable $z(x,t)=\varphi_\theta(h(x,t))$, using the same scalar
function at every spatial location. In latent space, the dynamics is modeled
as a linear stochastic evolution with parameters $K_\theta$ and $M_\theta$.
The decoder $\Psi_\theta$ maps the latent variable back to the physical field.
The locality of the transformation makes it directly interpretable as a
change of variables, while the linear latent dynamics provides a simple
description of the evolution.
} \label{fig:hero-schematic} 
\end{figure}

The structure of the model is illustrated in Fig.~\ref{fig:hero-schematic}.
We consider a dataset of stochastic interface evolutions, given by height fields
$h^t = (h_1^t, \dots, h_N^t)$ sampled at discrete times $t$, and denote by
$dh^t = h^{t+dt} - h^t$ the observed increment over a short time interval $dt$.
Our goal is to infer, directly from these observations, a dynamical model that
can reproduce the stochastic evolution of the system.

Motivated by the idea that complex stochastic dynamics may become simple in a
suitable representation, we seek a change of variables in which the evolution
takes a linear form. Specifically, we look for a transformation of the form
\[
z_i^t = \varphi_\theta(h_i^t), \qquad i=1,\dots,N,
\]
where the same scalar function $\varphi_\theta$ is applied independently at each
spatial site. This locality assumption is deliberately restrictive: the transformed
variable at site $i$ depends only on the height at the same site. The motivation is
twofold. First, such a transformation is directly interpretable, as it reduces to
a one-dimensional function that can be inspected a posteriori. Second, several
stochastic growth equations of physical interest admit local changes of variables
that simplify their dynamics, the Cole--Hopf transform being a notable example \cite{Corwin2012review}.

To ensure that the transformation is invertible, we simultaneously learn a map
\[
h_i^t = \Psi_\theta(z_i^t),
\]
and penalize reconstruction errors. The pair $(\varphi_\theta,\Psi_\theta)$
therefore defines a learned change of variables between the observed field $h$
and a new representation $z$.

In the transformed variables, we assume that the dynamics takes a simple stochastic
form. Specifically, we model the increments of $z$ as
\begin{equation}
    dz_i^t = \sum_j K_{\theta,ij}\, z_j^t\, dt
\;+\;
\sum_j M_{\theta,ij}\, z_j^t\, dW_j^t,
\label{eq:latent_sde_model}
\end{equation}
where $K_\theta$ and $M_\theta$ are $N\times N$ matrices, and $dW_j^t$ are independent
Gaussian increments with zero mean and variance $dt$. Equivalently, one can write
\[
dW_j^t = \sqrt{dt}\,\xi_j^t,
\]
where $\xi_j^t \sim \mathcal{N}(0,1)$ are independent standard normal variables.

The first term represents a linear deterministic evolution, while the second term
introduces multiplicative noise: each component $z_j^t$ sets the amplitude of an
independent noise source, which is then linearly mixed across space by the matrix
$M_\theta$. \rev{The learnt process noise is therefore zero-mean Gaussian
with covariance $M_\theta \operatorname{diag}(z^t)^2 M_\theta^\top$. This matrix is
symmetric and positive semi-definite by construction, for any $M_\theta$.}

The goal of the model is therefore to discover, directly from data, a representation
in which the dynamics closes in this linear form. \rev{We chose this parametrization for consistency with the linear Cole-Hopf dynamics (Eq. \ref{eq:z_lattice_sde}). More general forms generate the data just as well, but make the analytical map harder to recover unambiguously. We therefore report only this specific case.}

\subsection{Moment-matching loss}
\label{sec:loss}

The model is trained by matching the short-time statistics of the observed
increments. The reconstruction loss enforces consistency of the learned
change of variables:
\begin{equation}
    \mathcal{L}_{\rm rec}
    =
    \sum_i
    \left(
        \Psi_\theta(\varphi_\theta(h_i^t)) - h_i^t
    \right)^2 .
    \label{eq:loss_rec}
\end{equation}

The dynamical loss should enforce the agreement between dynamical variables, that is increments for a SDE. Since it is meaningless to ask for a full consistency between true and modeled trajectories, the loss is of a statistical kind. It is based on the conditional drift and covariance of the
increment predicted by the latent stochastic model. Since
$h_i^t=\Psi_\theta(z_i^t)$, It\^o's formula gives, to leading order in $dt$,
\begin{equation}
    dh_i^{t,\mathrm{model}}
    =
    \Psi_\theta'(z_i^t)\,dz_i^t
    +
    \frac{1}{2}\Psi_\theta''(z_i^t)(dz_i^t)^2 .
    \label{eq:ito_decoder}
\end{equation}
\rev{We denote by $dh^{t,\mathrm{model}}$ the increment
predicted by the model, and by $dh^{t,\mathrm{data}}$ the increment
measured in the training data.}
Substituting Eq.~\eqref{eq:latent_sde_model} yields the conditional drift
\begin{equation}
\begin{aligned}
    b_{\theta,i}(h^t)
    &:=
    \frac{\mathbb{E}[dh_i^{t,\mathrm{model}}\mid h^t]}{dt} \\
    &=
    \Psi_\theta'(z_i^t)
    \sum_j K_{\theta,ij}z_j^t
    +
    \frac{1}{2}\Psi_\theta''(z_i^t)
    \sum_j M_{\theta,ij}^2 (z_j^t)^2 .
\end{aligned}
\label{eq:model_drift}
\end{equation}
The second term is the It\^o correction induced by the multiplicative latent
noise and the nonlinear decoder.
The corresponding conditional covariance per unit time is
\begin{equation}
\begin{aligned}
    C_{\theta,ik}(h^t)
    &:=
    \frac{
    \mathrm{Cov}(dh_i^{t,\mathrm{model}},dh_k^{t,\mathrm{model}}\mid h^t)
    }{dt}
    \\
    &=
    \Psi_\theta'(z_i^t)\Psi_\theta'(z_k^t)
    \sum_j
    M_{\theta,ij}M_{\theta,kj}(z_j^t)^2 .
\end{aligned}
\label{eq:model_covariance}
\end{equation}

The empirical increment $dh^t$ is a noisy realization of the local transition
law. Therefore, the objective is not to match each observed increment exactly,
but to match the drift and covariance in expectation over many short-time
transitions. We use the empirical losses
\begin{equation}
    \mathcal{L}_{\rm drift}
    =
    \sum_i
    \left(
        b_{\theta,i}(h^t)
        -
        \frac{dh_i^{t,\mathrm{data}}}{dt}
    \right)^2 ,
    \label{eq:loss_drift}
\end{equation}
and
\begin{equation}
    \mathcal{L}_{\rm corr}
    =
    \sum_{i,k}
    \left(
        C_{\theta,ik}(h^t)
        -
        \frac{
        dh_i^{t,\mathrm{data}}
        dh_k^{t,\mathrm{data}}
        }{dt}
    \right)^2 .
    \label{eq:loss_corr}
\end{equation}
In Eq.~\eqref{eq:loss_corr} we use the raw product
$dh_i^{t,\mathrm{data}}dh_k^{t,\mathrm{data}}$ rather than the centered product.
This is equivalent at leading order in $dt$.  This correction is
therefore beyond the accuracy of the short-time expansion.

The total loss is
\begin{equation}
    \mathcal{L}
    =
    w_{\rm rec}\mathcal{L}_{\rm rec}
    +
    w_{\rm drift}\mathcal{L}_{\rm drift}
    +
    w_{\rm corr}\mathcal{L}_{\rm corr},
    \label{eq:total_physical_loss}
\end{equation}

with $w_{rec} = w_{drift} = w_{corr}=1$. This moment-matching perspective is close in spirit to Kramers--Moyal approaches, which estimate drift and diffusion from short-time increments \cite{FriedrichPeinke1997,Siegert1998,Boninsegna2018}. It differs from many latent-SDE methods, which typically rely on likelihood-based objectives over ensembles of trajectories and require rollouts \rev{(i.e. simulating the process)} during training \cite{Heck2025noise}. Here, instead, we use a local-in-time moment-matching principle in a learned latent representation, while imposing a physically interpretable structure: a local scalar change of variables and a linear stochastic latent dynamics.

\subsection{Implementation and training}

The model consists of four components: the encoder $\varphi_\theta$, the decoder
$\Psi_\theta$, and the matrices $K_\theta$ and $M_\theta$ defining the latent dynamics.
All components are trained jointly. The encoder and decoder are scalar feed-forward neural networks with three hidden
layers of 128 units and smooth (SiLU) activations. Because they act independently
at each spatial site, their evaluation and differentiation remain computationally cheap.
The matrices $K_\theta$ and $M_\theta$ are treated as free $N \times N$ parameters,
with no spatial structure imposed a priori. In particular, no Laplacian or locality
constraint is enforced. The loss function depends on the first and second derivatives of $\Psi_\theta$,
which are computed by automatic differentiation, meaning that the derivatives
are obtained analytically by propagating gradients through the layers of the
network, rather than by numerical finite differences. This remains efficient since
$\Psi_\theta$ is one-dimensional.

Training is performed using the Adam optimizer with learning rate $10^{-5}$ and $L_2$ regularization on the encoder and decoder parameters. Each training step
uses a batch of independently sampled transitions $(h^t, dh^t)$.
Importantly, the loss is evaluated on single-step transitions and does not require
simulating trajectories during training. This makes the procedure significantly
cheaper than likelihood-based latent-SDE approaches.

\begin{figure}[h]
\centering
\includegraphics[width=0.5\columnwidth]{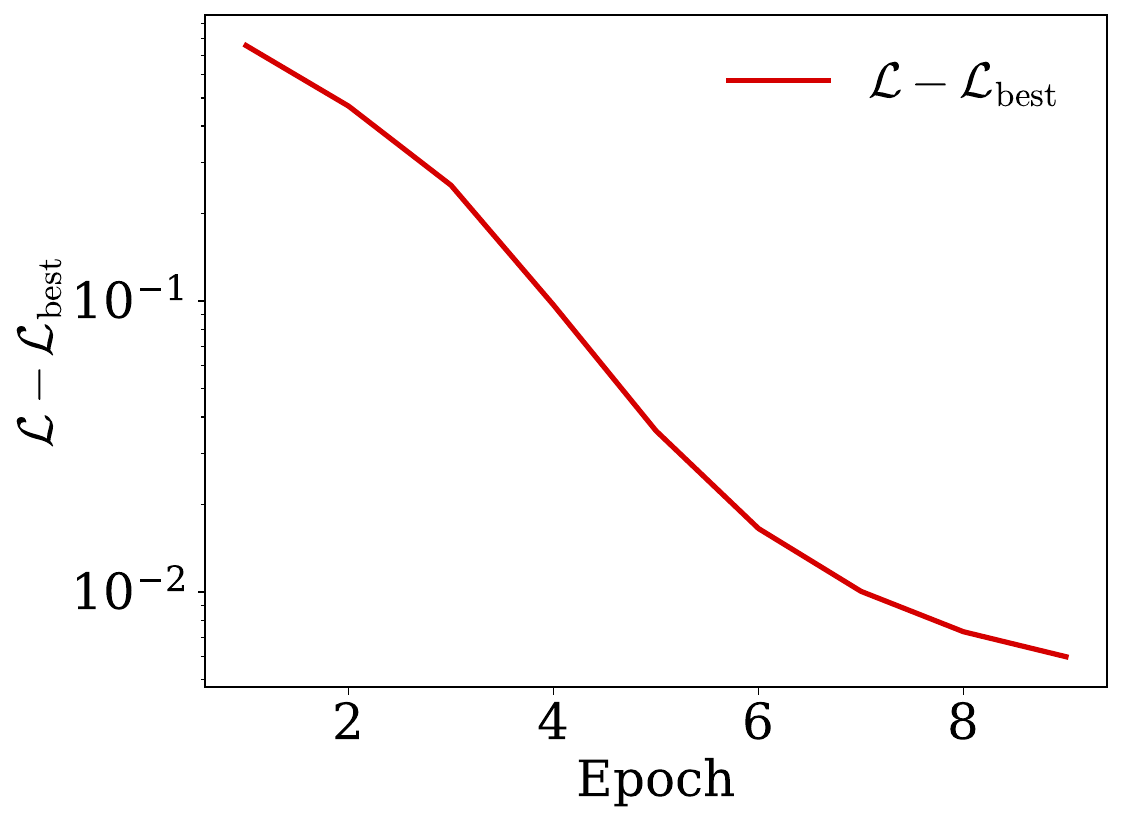}
\caption{
Training dynamics.
The difference between the total loss $\mathcal{L}$ and the optimal loss $\mathcal{L}_{best}$  is shown as a function of the training epoch.
$\mathcal{L}_{best}$ is computed by evaluating the loss at the exact Cole--Hopf solution. Due to noise fluctuations, $\mathcal{L}_{best}$ is not zero.  
}
\label{fig:convergence}
\end{figure}
As a validation check, we periodically generate trajectories by encoding an initial
condition, evolving the latent dynamics forward in time, and decoding back to the
observation space. We monitor the agreement between generated \rev{data} and empirical
distributions \rev{of the training data} using the symmetric Kullback--Leibler divergence. 
Training is stopped
when this metric saturates.

The model is trained on this dataset using the loss described in section \ref{sec:loss}.
Figure~\ref{fig:convergence} shows the evolution of the loss during training.
A notable feature is that the loss does not vanish, and should not.
This reflects the stochastic nature of the data: the observed increments
$dh^t$ contain an irreducible stochastic contribution, so an exact match
between model predictions and observations is not expected.

\subsection{Generative accuracy}
\label{subsec:physical_generative_accuracy}

\begin{figure}[h]
\centering
\subfloat[\label{fig:kpz-mean}]{
\includegraphics[width=0.45\textwidth]{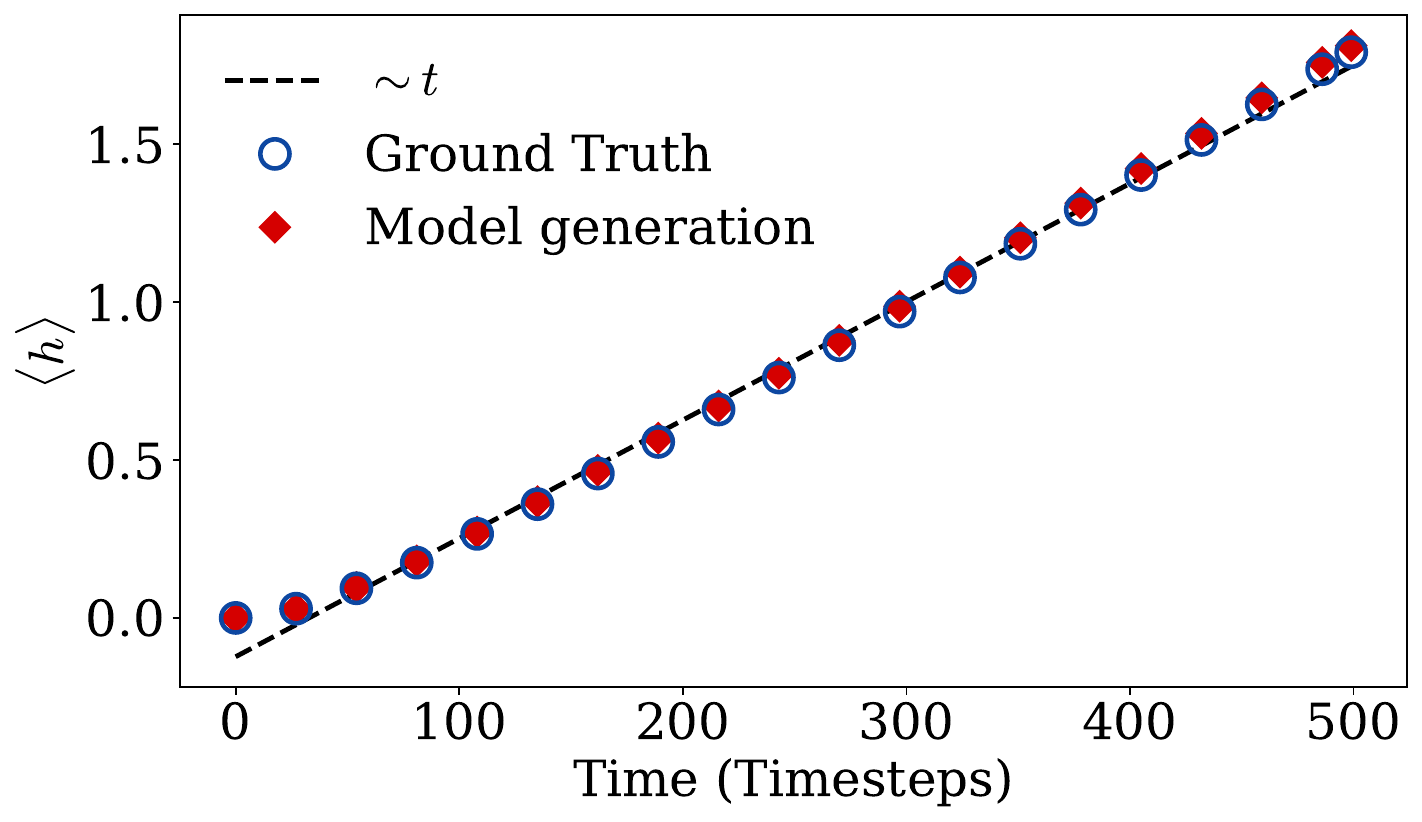}}
\hfill
\subfloat[\label{fig:kpz-var}]{
\includegraphics[width=0.45\textwidth]{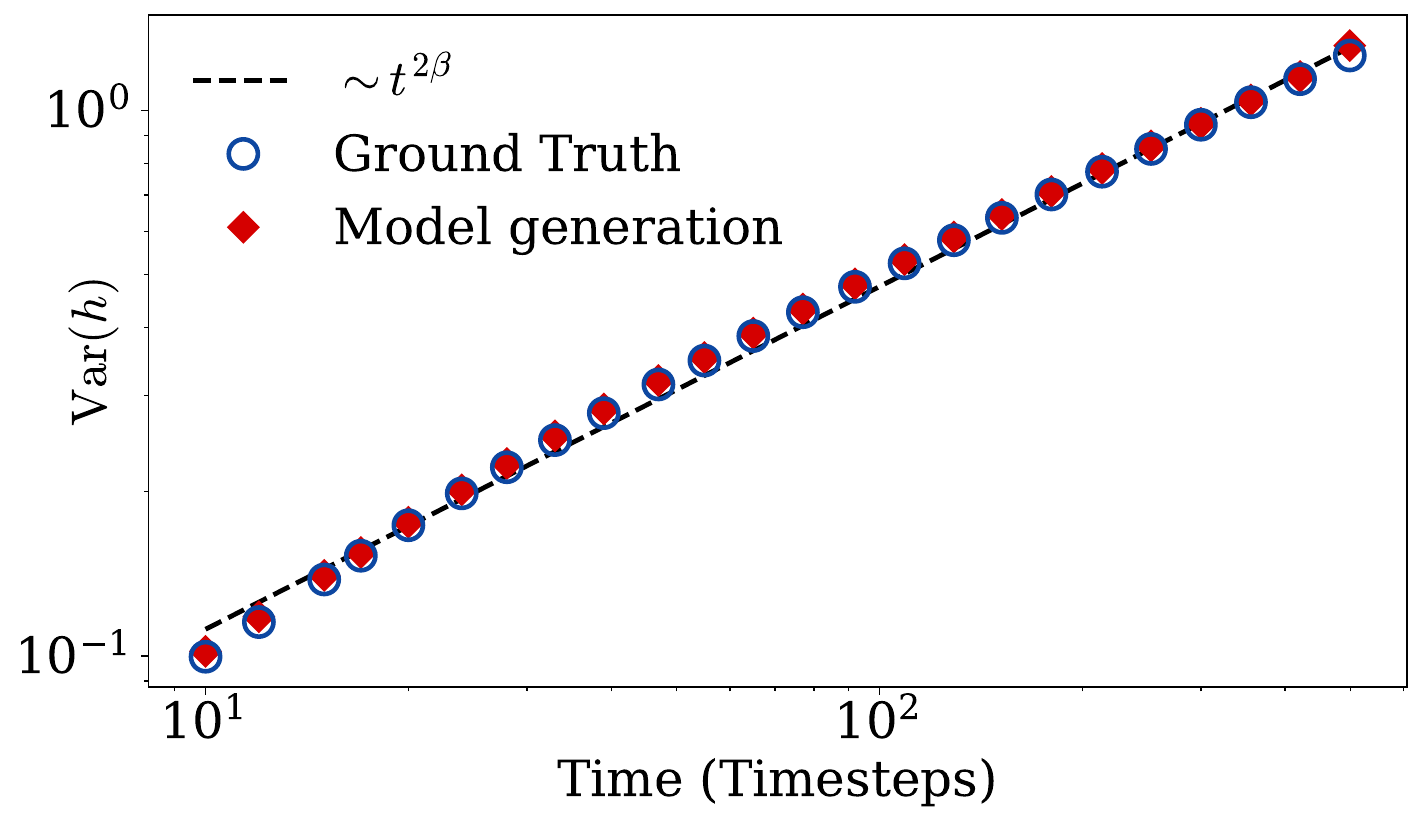}}\\[6pt]
\subfloat[\label{fig:kpz-tw-lin}]{
\includegraphics[width=0.45\textwidth]{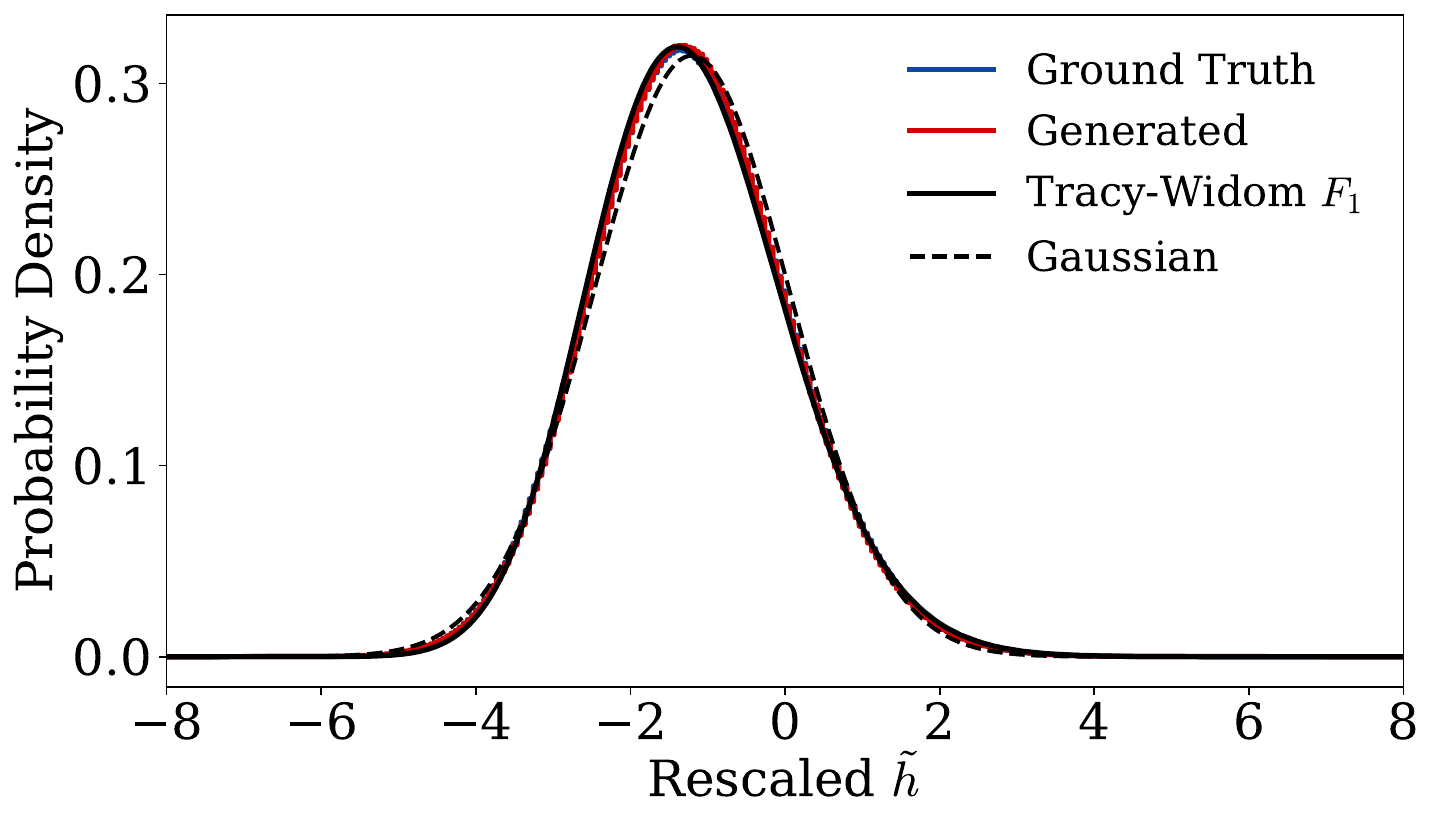}}
\hfill
\subfloat[\label{fig:kpz-tw-log}]{
\includegraphics[width=0.45\textwidth]{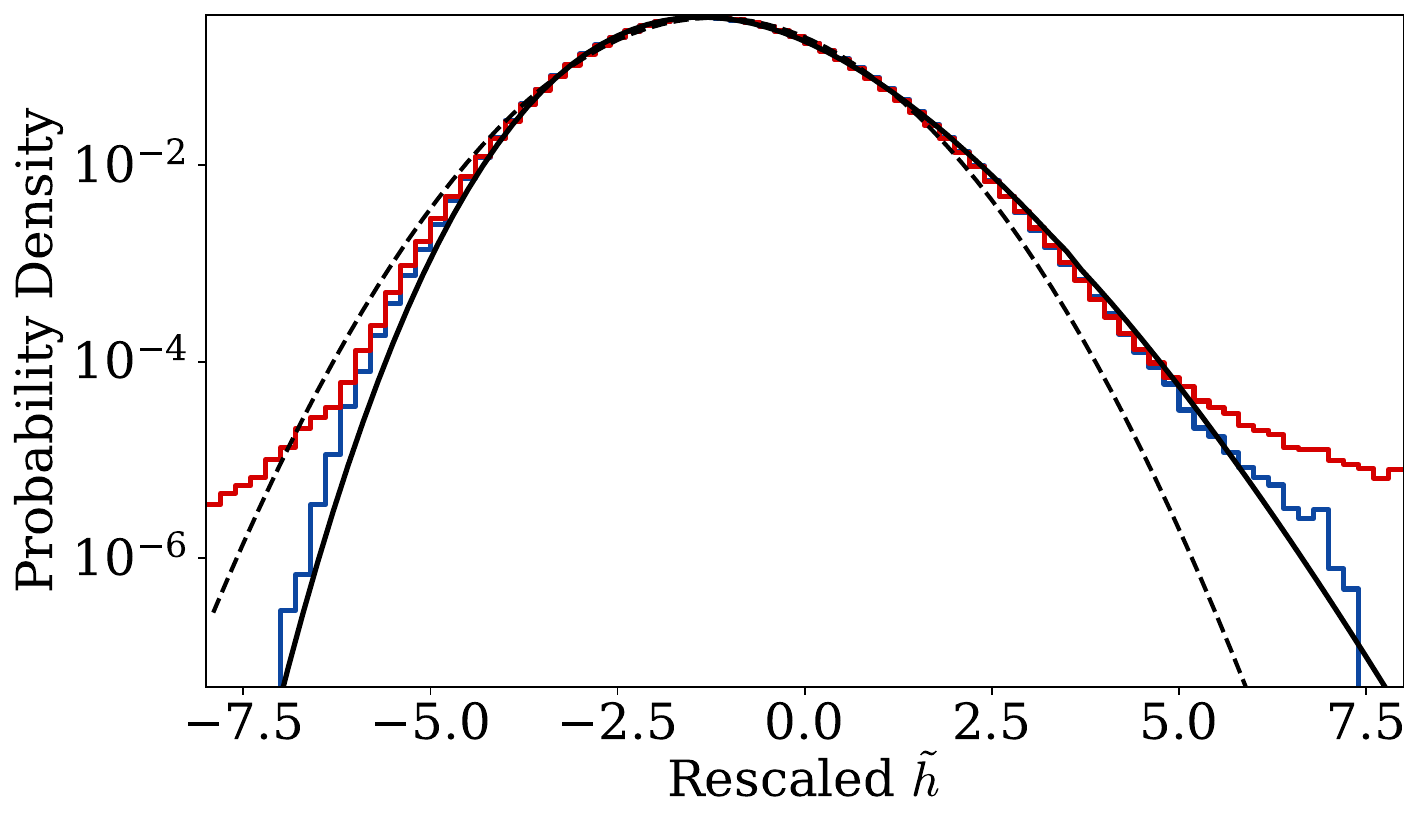}}
\caption{
KPZ statistics: generated model trajectories versus ground truth.
(a) Ensemble-averaged mean height as a function of time.
(b) Height variance, displaying the expected KPZ scaling
$\mathrm{Var}(h)\sim t^{2/3}$.
(c,d) Distribution of the rescaled height $\widetilde h$ at late times, shown
on linear and semi-logarithmic scales. The theoretical reference is the GOE
Tracy--Widom law associated with flat initial conditions, \rev{which exhibit an asymmetric non-Gaussian decay. We attribute the tail discrepancies to small deviations of the learned components from the true solution.}
}
\label{fig:kpz}
\end{figure}

We evaluate the learned model in two complementary ways. First, we test its
generative accuracy by comparing trajectories sampled from the model with
the empirical KPZ dataset. Second, we inspect the learned transformation and
latent operators to determine whether the model has recovered the underlying
physical structure.

The first diagnostic is the ensemble-averaged mean height, which grows
linearly in time,
\begin{equation}
    \langle h^{\rev{t}}\rangle \sim v_\infty t .
\end{equation}
The second is the height variance, which in the one-dimensional KPZ class
grows as
\begin{equation}
    \mathrm{Var}(h^{\rev{t}})\sim t^{2\beta},
    \qquad
    \beta=\frac{1}{3}.
    \label{Eq:variance}
\end{equation}
Finally, we compare the full one-point height distribution. For flat initial
conditions, the rescaled height converges in the asymptotic KPZ regime to the
GOE Tracy--Widom distribution. We use
\begin{equation}
    \widetilde h
    =
    \sigma_{\rm TW}
    \frac{h-\langle h\rangle}{\sqrt{\mathrm{Var}(h)}}
    +
    \mu_{\rm TW},
    \label{eq:TW_rescaling}
\end{equation}
\rev{where $\mu_{TW}=-1.2065$ and $\sigma_{TW}=1.2680$ are the mean and
standard deviation of the standard GOE Tracy--Widom distribution.
For flat KPZ growth one  commonly uses the convention 
$\sigma_{\rm flat}=2^{-2/3}\sigma_{TW}$ \cite{PrahoferSpohn2000}. This constant factor does not affect the shape of the distribution and here we use the standard GOE convention only to fix the axes of the plots.}

Figure~\ref{fig:kpz} shows that the learned model reproduces these KPZ
statistics. \rev{The trend lines displayed in Fig.\ref{fig:kpz-mean} and  Fig. \ref{fig:kpz-var} are obtained as least squared regression on the Ground Truth points.} The generated trajectories exhibit the correct linear growth of
the mean, the expected roughening exponent, and a late-time rescaled height
distribution close to the GOE Tracy--Widom law. This indicates that the model
does not merely fit short-time increments, but also generates trajectories
whose long-time statistics remain in the KPZ universality class.

\subsection{Recovery of the Cole--Hopf structure}
\label{subsec:recovery_cole_hopf}

\begin{figure}[h]
\centering
\includegraphics[width=0.5\columnwidth]{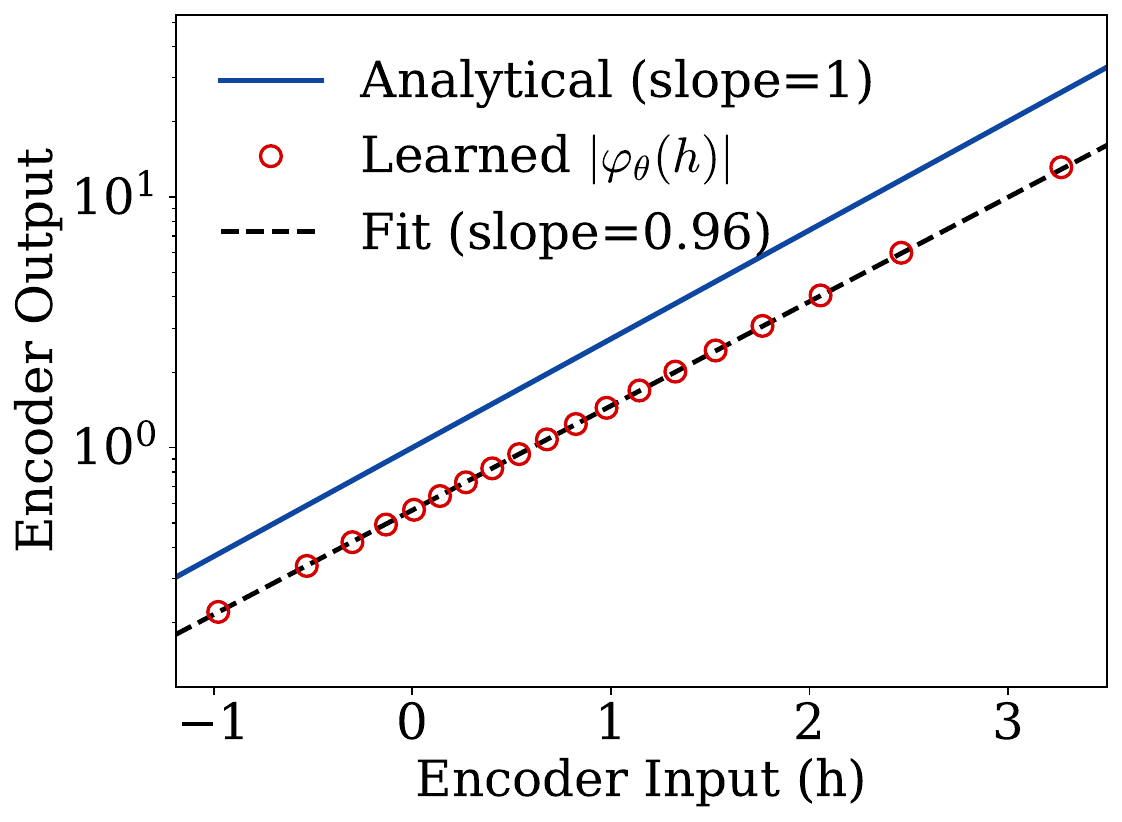}
\caption{
Learned change of variables. The learned mapping
$|\varphi_\theta(h)|$ follows an exponential dependence on $h$, in agreement
with the \rev{analytical} Cole-Hopf transform $z=e^{\alpha h}$.
The vertical offset corresponds to an arbitrary multiplicative constant in
the latent variable, which does not affect the linear latent dynamics. 
}
\label{fig:encoder}
\end{figure}

We finally ask whether the model has identified the physical representation
underlying the KPZ dynamics. For KPZ, this amounts to recovering the
Cole--Hopf transformation and the associated stochastic heat equation.

Figure~\ref{fig:encoder} shows the learned scalar map $\varphi_\theta$.
In semi-logarithmic scale, the learned function is approximately linear,
indicating an exponential dependence on $h$. A fit gives
\begin{equation}
    \varphi_\theta(h) \simeq C e^{\widehat{\alpha}h},
    \qquad
    \widehat{\alpha}\simeq 0.96,
\end{equation}
close to the theoretical value $\alpha=\lambda/(2\nu)=1$ used in the
simulations. The multiplicative constant $C$ is not identifiable, since
rescaling $z$ by a constant leaves the linear latent dynamics unchanged after
a corresponding rescaling of the decoder. Up to this trivial degree of
freedom, the model therefore recovers the Cole--Hopf map.

\begin{figure}[h]
\centering
\includegraphics[width=0.98\textwidth]{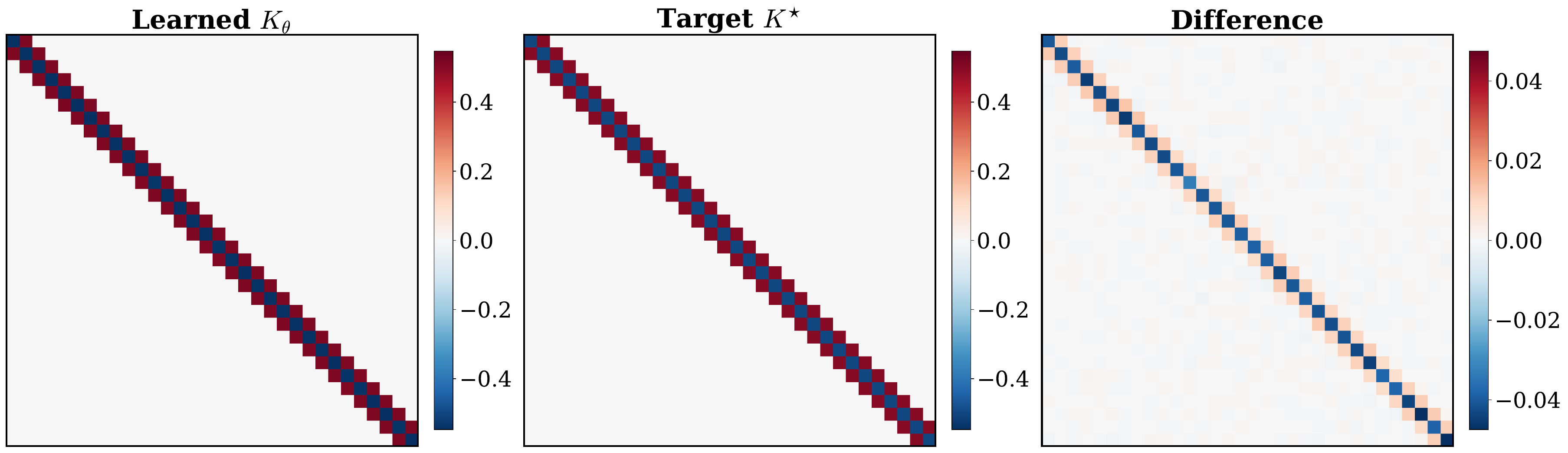}
\caption{
Learned drift matrix $K_\theta$ compared with the analytical target
$K^\star$. Left: learned matrix. Center: analytical discrete diffusion
operator. Right: difference $K_\theta-K^\star$. The first $32\times 32$ block
is shown. The learned operator recovers the tridiagonal structure of the
discrete Laplacian and the periodic boundary conditions, although small
residual deviations remain.
}
\label{fig:matrixK}
\end{figure}
Figure~\ref{fig:matrixK} compares the learned drift matrix $K_\theta$ with
the analytical operator $K^\star$ corresponding to the discrete stochastic
heat equation. The dominant tridiagonal structure of the discrete Laplacian is
recovered without imposing any stencil or locality constraint. The learned
matrix also captures the periodic boundary conditions. Small deviations from
the ideal operator remain and are visible in the difference plot. These errors
are consistent with the small discrepancies observed in the far tails of the
height distribution, where rare fluctuations are particularly sensitive to the
precise latent noise structure.
\rev{We also show the learnt matrix $M_{\theta}$ in Appendix \ref{app:M}. The model autonomously recovers the diagonal structure of the ground truth.}

Taken together, these results show that the model does not only reproduce
KPZ statistics. It also reconstructs explicit, interpretable objects: a
local exponential change of variables and a linear latent stochastic operator.
These objects are sufficient to infer the underlying stochastic heat equation
and, by inversion through the learned decoder, the corresponding KPZ dynamics.

\rev{\section{Proof-of-concept extension to a flexible autoencoder}}
\label{subsec:flexible_autoencoder_generalisation}

The interpretable latent-SDE model introduced above is deliberately constrained:
the change of variables is local and scalar, and the latent dynamics is linear.
These assumptions make the learned representation directly readable, and allow
us to match the short-time drift and covariance of the observed increments. 
\rev{In this last section, we provide some insights on the question whether the same statistical-learning
philosophy could also be useful for a less constrained architecture.}

As a proof of concept, we consider a convolutional autoencoder architecture
similar in spirit to those used in deterministic latent-linearization
approaches~\cite{Lusch2018}. In contrast with the local model of
Sec.~\ref{subsec_arch}, the encoder and decoder are now general
convolutional maps. The latent representation is therefore no longer restricted
to be a pointwise change of variables, and the learned coordinates cannot be
directly compared with the scalar Cole--Hopf transform. This makes the model
less interpretable, but potentially more flexible for systems where the
simplifying representation is nonlocal or not known in advance.  \rev{One could extend the loss formulation of section \ref{sec:loss} to the multivariate case by using the Itô chain rule for multivariate transformations. This, however, would require the computation of the Hessian matrix of the transform. For discretized spatially extended systems (in our case, 256 sites), this can become computationally expensive. We therefore propose another learning strategy.  }

In the interpretable model, the latent dynamics has a simple explicit form,
which allows us to match the local short-time drift and covariance of the
observed increments. For the flexible autoencoder, our goal is different: we
do not attempt to extract a readable stochastic equation, but only to test
whether statistical constraints improve the quality of generated trajectories.
We therefore implement the same learning philosophy at the trajectory level:
individual stochastic trajectories are not matched pointwise, while ensemble
observables of the generated trajectories are constrained to agree with those
of the KPZ dataset.

The model is trained by generating trajectories from the latent dynamics and
matching their statistics to those of the KPZ dataset. \rev{We keep the
notation of Sec.~\ref{sec:loss}: the superscripts $\mathrm{model}$ and
$\mathrm{data}$ distinguish quantities evaluated on the generated
trajectories from those measured on the training set, while
$\langle\cdot\rangle$ denotes the average over the $N$ lattice sites and over
the independent noise realizations of the batch. The index $t=1,\dots,T$ runs
over the time steps of the rollout.}
 
In addition to the reconstruction loss,
\begin{equation}\revc
\mathcal{L}_{\rm rec}
=
\frac{1}{T+1}
\sum_{t=0}^{T}
\left\langle
\left(
\Psi_\theta\!\left(\varphi_\theta(h^{t,\mathrm{data}})\right)
-
h^{t,\mathrm{data}}
\right)^{2}
\right\rangle ,
\label{eq:flexible_rec_loss}
\end{equation}
\rev{which has the same meaning as Eq.~\eqref{eq:loss_rec},} we impose
agreement between generated and empirical observables. The first two terms
match the mean height and the \rev{height variance already used in
Fig.~\ref{fig:kpz}},
\begin{equation}\revc
\mathcal{L}_{\rm mean}
=
\frac{1}{T}
\sum_{t=1}^{T}
\left(
\left\langle h^{t,\mathrm{model}} \right\rangle
-
\left\langle h^{t,\mathrm{data}} \right\rangle
\right)^{2} ,
\label{eq:flexible_mean_loss}
\end{equation}
and
\begin{equation}\revc
\mathcal{L}_{\rm var}
=
\frac{1}{T}
\sum_{t=1}^{T}
\left(
\mathrm{Var}\!\left(h^{t,\mathrm{model}}\right)
-
\mathrm{Var}\!\left(h^{t,\mathrm{data}}\right)
\right)^{2} ,
\label{eq:flexible_var_loss}
\end{equation}
\rev{where $\mathrm{Var}(h^{t})$ is the height variance, defined in Eq. \ref{Eq:variance}.}
 
We also include a spatial-correlation loss. For a trajectory ensemble, define
\begin{equation}\revc
G^{t}(\xi)
=
\left\langle
\left(
h_i^{t}-\left\langle h^{t}\right\rangle
\right)
\left(
h_{i+\xi}^{t}-\left\langle h^{t}\right\rangle
\right)
\right\rangle ,
\label{eq:spatial_correlation}
\end{equation}
with periodic indexing, \rev{so that $G^{t}(0)=\mathrm{Var}(h^{t})$.} The
correlation loss is
\begin{equation}\revc
\mathcal{L}_{G}
=
\frac{1}{T|\Xi|}
\sum_{t=1}^{T}
\sum_{\xi\in\Xi}
\left(
G^{t,\mathrm{model}}(\xi)-G^{t,\mathrm{data}}(\xi)
\right)^{2} ,
\label{eq:flexible_corr_loss}
\end{equation}
where $\Xi$ is the set of spatial lags used during training. The total loss is
therefore
\begin{equation}
\mathcal{L}_{\rm flex}
=
w_{\rm rec}\mathcal{L}_{\rm rec}
+
w_{\rm mean}\mathcal{L}_{\rm mean}
+
\rev{w_{\rm var}\mathcal{L}_{\rm var}}
+
\rev{w_{G}\mathcal{L}_{G}},
\label{eq:flexible_total_loss}
\end{equation}
with
\begin{equation}
    w_{\rm rec}=0.1,
    \qquad
    w_{\rm mean}=1,
    \qquad
    \rev{w_{\rm var}=1},
    \qquad
    \rev{w_{G}=1} .%
\end{equation}
The training dataset for this setup was generated using $\lambda=2$ and
$dt=0.05$.

\begin{figure}[h]
\centering
\subfloat[\label{fig:mean}]{
\includegraphics[width=0.45\textwidth]{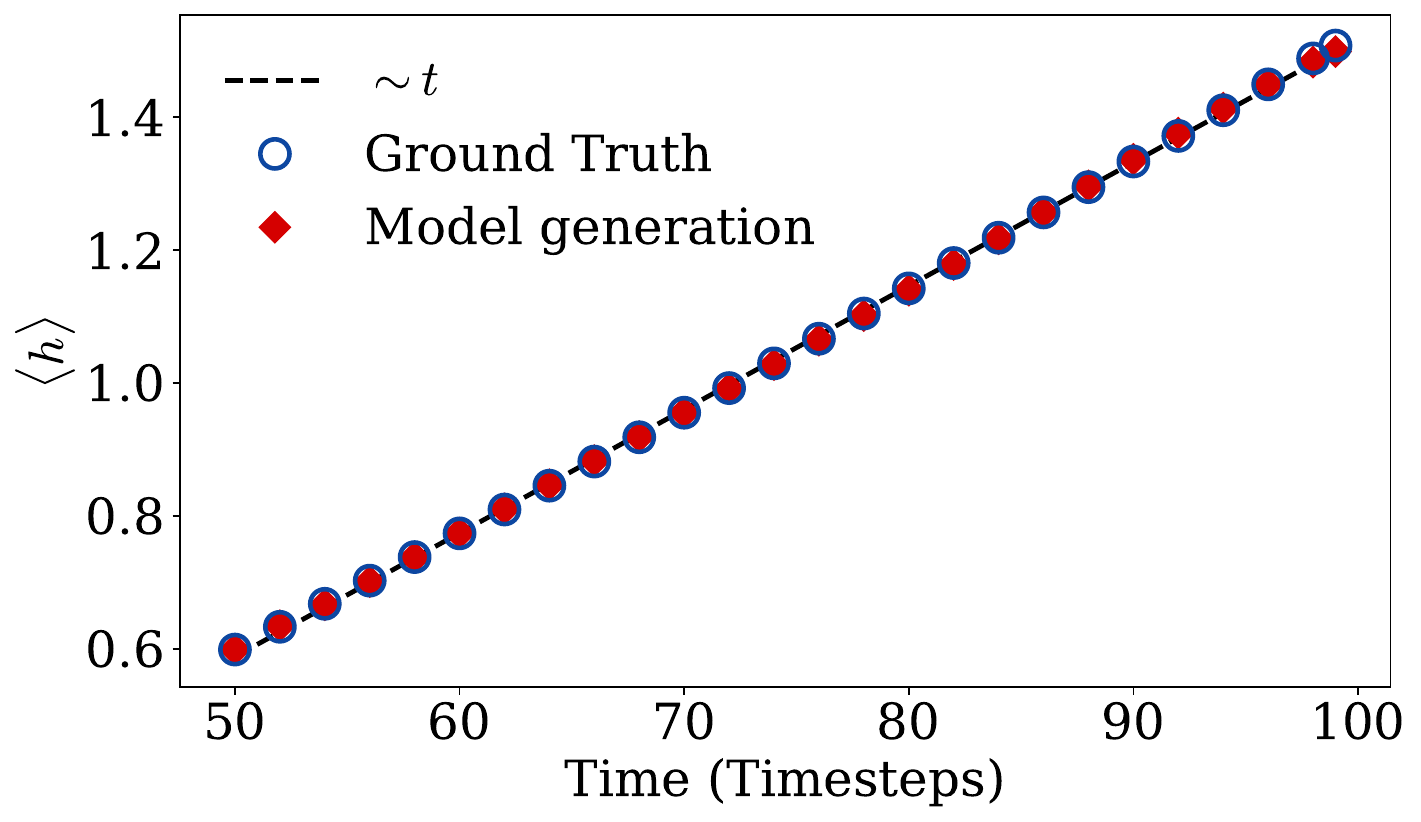}}
\hfill
\subfloat[\label{fig:variance}]{
\includegraphics[width=0.45\textwidth]{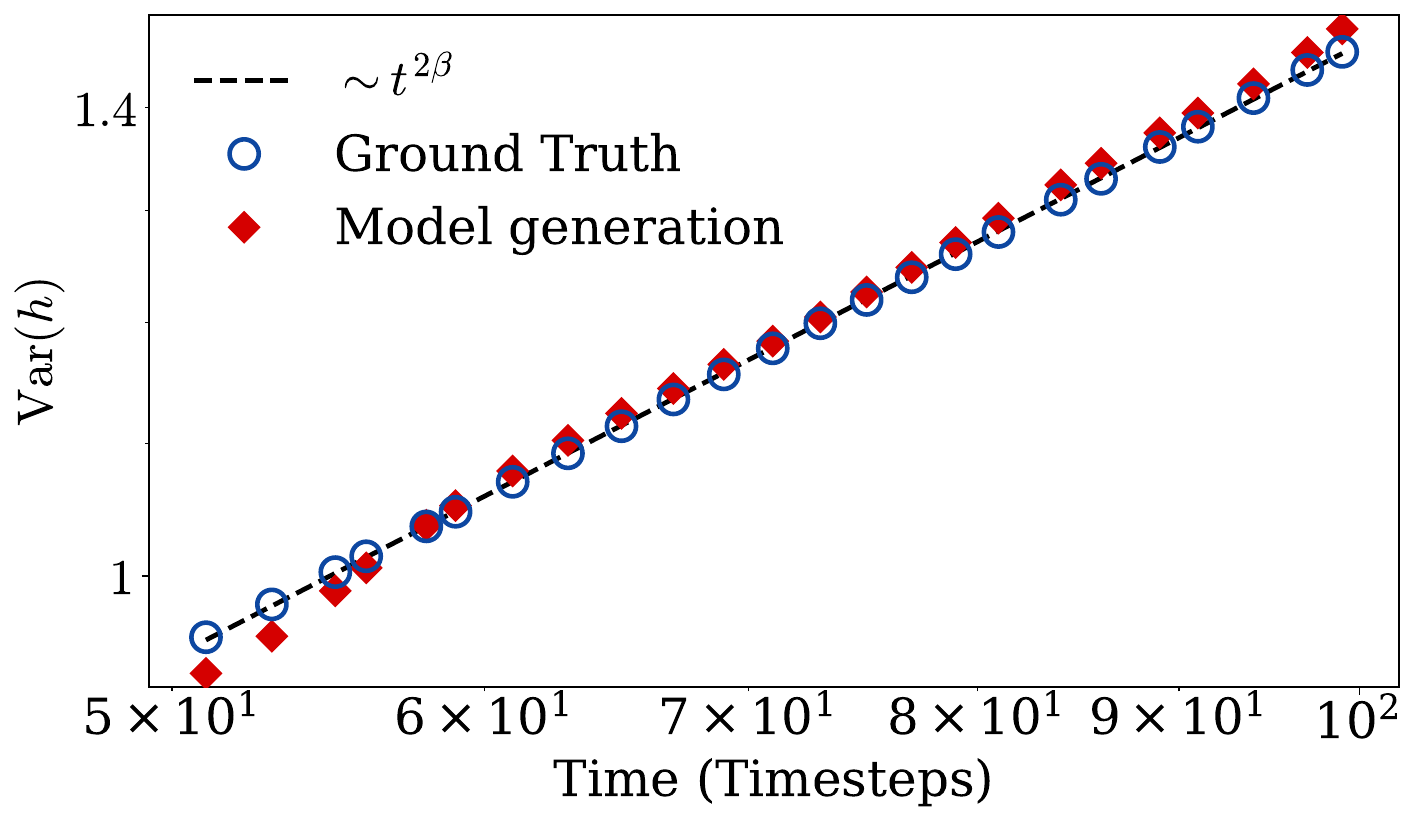}}
\caption{
KPZ statistics: model trajectories versus ground truth. The flexible autoencoder was used for these plots.  (a) Mean height and (b) height variance.   Dashed lines show the KPZ scalings
$\langle h\rangle\sim t$ and $\mathrm{Var}(h)\sim t^{2\beta}$. 
}
\label{fig:mean_variance}
\end{figure}

Figure~\ref{fig:mean_variance} compares the mean height
\rev{$\langle h^{t}\rangle$} and the \rev{height variance
$\mathrm{Var}(h^{t})$} for the ground-truth KPZ trajectories and for
trajectories generated by the trained autoencoder. The model captures the
linear growth of the mean height and the roughening of the interface, with
residual discrepancies at short times and near the end of the rollout.

\begin{figure}[t]
\centering
\includegraphics[width=0.6\columnwidth]{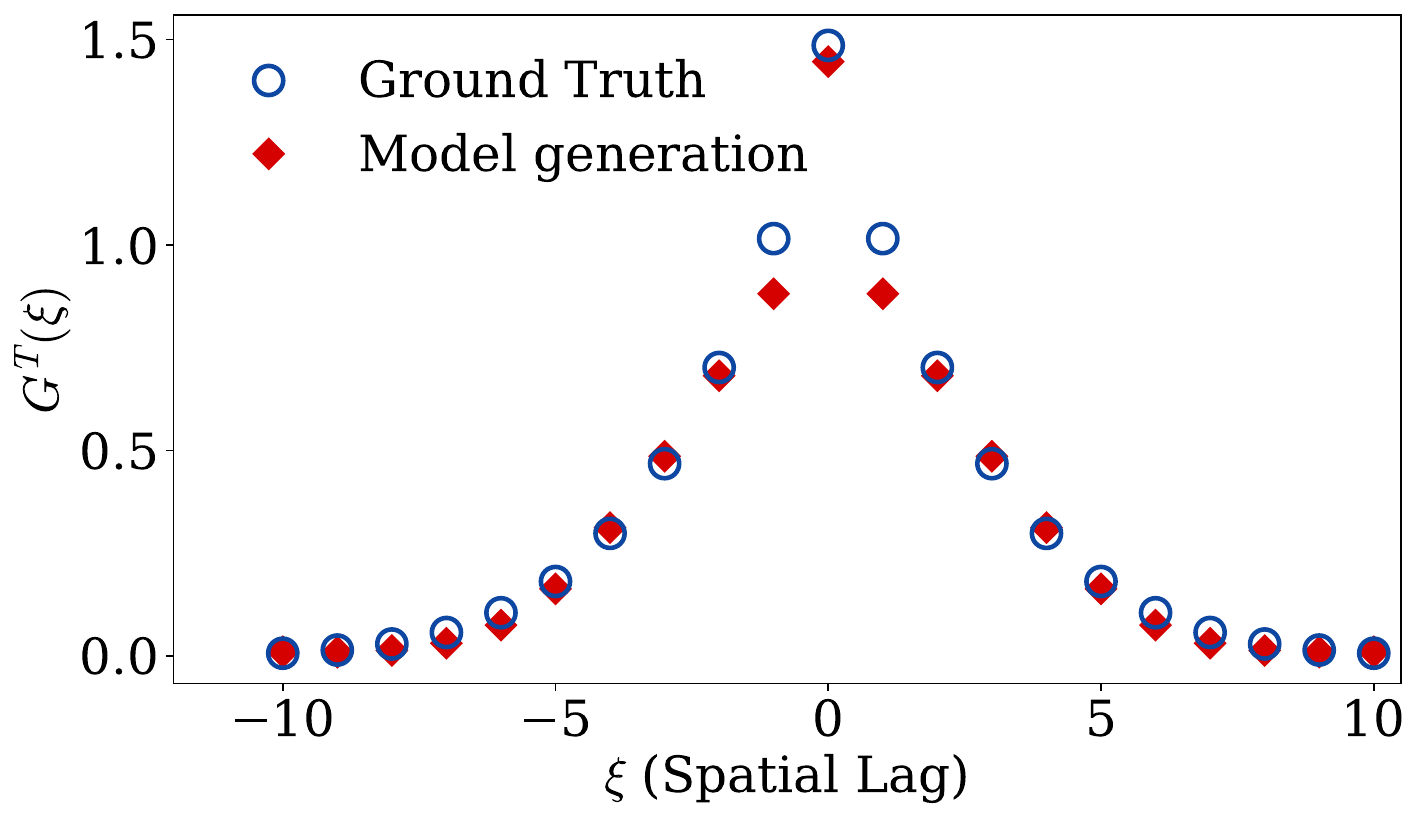}
\caption{
Spatial correlation function $G^T(\xi)$  (Eq. \ref{eq:spatial_correlation}) at final time.
The flexible autoencoder model reproduces the spatial correlation length of KPZ. 
}
\label{fig:spatial_corr_overlay}
\end{figure}

The spatial correlations provide a more stringent test than the first two
global moments. As shown in Fig.~\ref{fig:spatial_corr_overlay}, the generated
trajectories reproduce the shape and time evolution of the empirical
correlation function $G^t(\xi)$ with reasonable accuracy. This indicates that
the model has captured part of the spatial organization of the KPZ interface,
and not only its one-point statistics.

\rev{While this loss formulation is not directly comparable with the one defined in Sec. \ref{sec:loss}, we think it could be used as a "proof of concept" for more complex settings.} Unlike the
physically constrained model, the flexible autoencoder does not provide a
directly readable Cole--Hopf map or an explicit stochastic heat operator.
However, it shows that statistical losses based on ensemble observables can
improve the generative behavior of more expressive architectures. \rev{At this early stage, we show that one can encode in the loss relevant statistics of a nonlinear process and produce, in a linear way, new trajectories that follow these key statistics.} This suggests
that the moment-matching philosophy may be useful beyond the local
interpretable setting, especially for stochastic systems where the appropriate
simplifying variables are unknown or nonlocal.
\section{Conclusion}
\label{sec:conclusion}

We have introduced an interpretable latent-SDE framework for learning changes
of variables that transform nonlinear stochastic dynamics into a simple linear
stochastic evolution. The model combines a learned local transformation with a
linear latent dynamics and is trained through a moment-matching objective based
on short-time increments, avoiding trajectory rollouts during optimization.

We validated the approach on the one-dimensional KPZ equation, where the
Cole--Hopf transformation provides an exact reference solution. Starting from
raw trajectories alone, the model successfully recovers both the local
exponential change of variables and the associated linear stochastic
operator. The learned representation is therefore not only generative but also
interpretable, providing explicit mathematical objects from which the
underlying stochastic evolution can be reconstructed.

The generated trajectories reproduce the main statistical signatures of the
KPZ universality class, including the linear growth of the mean height, the
roughening law $\rev{\mathrm{Var}(h^{\rev{t}})}\sim t^{2\beta}$ with $\beta=1/3$, and the convergence
toward the GOE Tracy--Widom distribution for flat initial conditions. These
properties are not imposed during training, but emerge naturally from matching
the local statistics of the stochastic dynamics.

The present framework nevertheless has several limitations. It relies on
short-time increments and therefore requires data with sufficient temporal
resolution. Moreover, the loss constrains only one-step statistics, so small
errors may accumulate during long rollouts. Finally, the interpretable model
assumes a local scalar change of variables, an assumption well suited to KPZ
but potentially too restrictive for more general stochastic field theories.
\rev{Relatedly, the optimization always returns some transformation, so the loss alone does not certify that a linearizing representation exists. In practice this is assessed a posteriori: if the statistics of the generated trajectories reproduce those of the dataset the learned representation can be considered adequate. A sharper test is to monitor the generative quality while increasing the model expressivity and the size of the training set: a saturation of the performance at a finite error suggests that the mapping is only approximate rather than exact. Even in that case the result remains physically informative, since a simple effective model capturing part of the dynamics is a natural starting point for analytical work.}

As a proof of concept, we also investigated a more flexible convolutional
autoencoder trained according to the same statistical-learning philosophy,
although using trajectory-level observables instead of local increment
statistics. While this architecture is no longer interpretable and does not
recover the Cole--Hopf transformation explicitly, it reproduces several key
KPZ observables with reasonable accuracy. This suggests that statistical
moment-matching strategies may also prove useful for more expressive neural
architectures.

Future work will explore the application of this autoencoder framework to other equations without known linearizing transformations.

\appendix

\section{Architecture details for the flexible autoencoder}
\label{sec:black_box_autoencoder}

This appendix describes the implementation details of the proof-of-concept
experiment presented in Sec.~IV. Its purpose is to investigate whether the
statistical-learning philosophy developed for the interpretable latent-SDE
model can also improve a more expressive, but non-interpretable, neural
architecture.
\begin{figure*}[t]
    \centering
    \includegraphics[width=\textwidth]{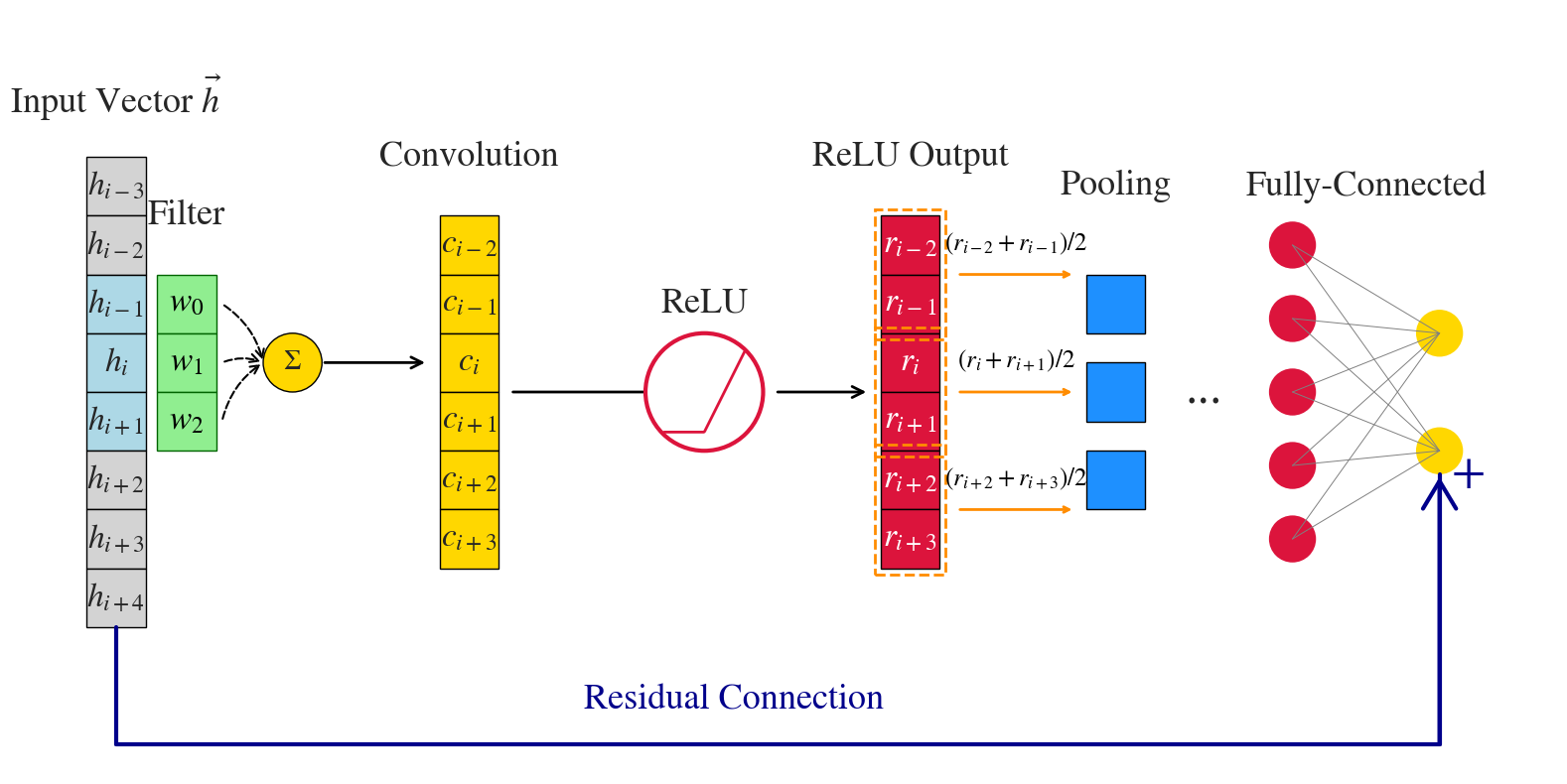}
    \caption{
    Schematic representation of the convolutional block used in the flexible
    autoencoder. A one-dimensional convolutional filter is applied locally to
    the input height profile, followed by a nonlinear activation and pooling.
    The number of feature maps is increased across successive convolutional
    layers. The final feature maps are flattened and passed through fully
    connected layers. A residual connection is added to the output, producing
    a dimension-preserving transformation from the height field $\mathbf{h}$
    to the latent field $\mathbf{z}$.
    }
    \label{fig:cnn}
\end{figure*}

 The encoder and decoder are implemented using
one-dimensional convolutional neural networks, a natural choice for spatially
extended systems because they exploit locality and weight
sharing~\cite{LeCun1998,cacciari2024hands} \rev{: each output depends only on a small neighborhood of the input, and the same filter is applied at every spatial position.} \rev{Convolutions are also a natural fit for physical systems, since a small kernel can represent local differential operators directly.} The architecture is inspired by
the deterministic latent-linearization framework of Ref.~\cite{gin2021deep}. 

Let
\begin{equation}\revc
    h^{t}
    =
    \left(h_1^{t},\dots,h_N^{t}\right)
\end{equation}
denote the discretized KPZ height profile \rev{at time $t$, as in
Sec.~\ref{subsec_arch}}. The encoder
\begin{equation}
    \rev{\varphi_\theta} : \mathbb{R}^{N}\to\mathbb{R}^{N}
\end{equation}
maps the height field to a latent field,
\begin{equation}\revc
    z^{t}=\varphi_\theta\!\left(h^{t}\right),
    \label{eq:black_box_encoder}
\end{equation}
while the decoder
\begin{equation}
    \rev{\Psi_\theta} : \mathbb{R}^{N}\to\mathbb{R}^{N}
\end{equation}
maps the latent field back to the height representation,
\begin{equation}\revc
    \Psi_\theta\!\left(z^{t}\right)
    =
    \Psi_\theta\!\left(\varphi_\theta\!\left(h^{t}\right)\right).
    \label{eq:black_box_decoder}
\end{equation}
\rev{Unlike in Sec.~\ref{subsec_arch}, $\varphi_\theta$ and $\Psi_\theta$ now
act on the whole profile rather than site by site, so the change of variables
is no longer local.}
As in the interpretable model, the latent dimension is chosen equal to the
number of spatial lattice sites. \rev{This choice, also tested in \cite{gin2021deep}, avoids the information loss imposed by a dimensional reduction and is compatible with an invertible change of variables.}
To generate trajectories, we initialize the latent field by encoding the flat
initial condition,
\begin{equation}\revc
    z^{0,\mathrm{model}}
    =
    \varphi_\theta\!\left(h^{0}\right),
    \qquad
    h^{0}=0,
    \label{eq:black_box_initial_condition}
\end{equation}
and evolve it with the same stochastic linear update as used in the other
model [Eq.~\eqref{eq:latent_sde_model}]. The resulting architecture is more
flexible, but no longer directly interpretable.
 
The generated height trajectories are finally obtained by decoding the latent
rollout,
\begin{equation}\revc
    h^{t,\mathrm{model}}
    =
    \Psi_\theta\!\left(z^{t,\mathrm{model}}\right).
    \label{eq:black_box_generated_height}
\end{equation}
Unlike the interpretable architecture, the purpose of this model is not to
recover the Cole--Hopf transformation itself, but to assess whether the
statistical-learning strategy developed in the main text remains beneficial \rev{with an architecture capable of representing more complex transforms}.

\section{The learned coupling matrix $M$}
\label{app:M}

\begin{figure*}[t]
  \centering
  \includegraphics[width=0.9\linewidth]{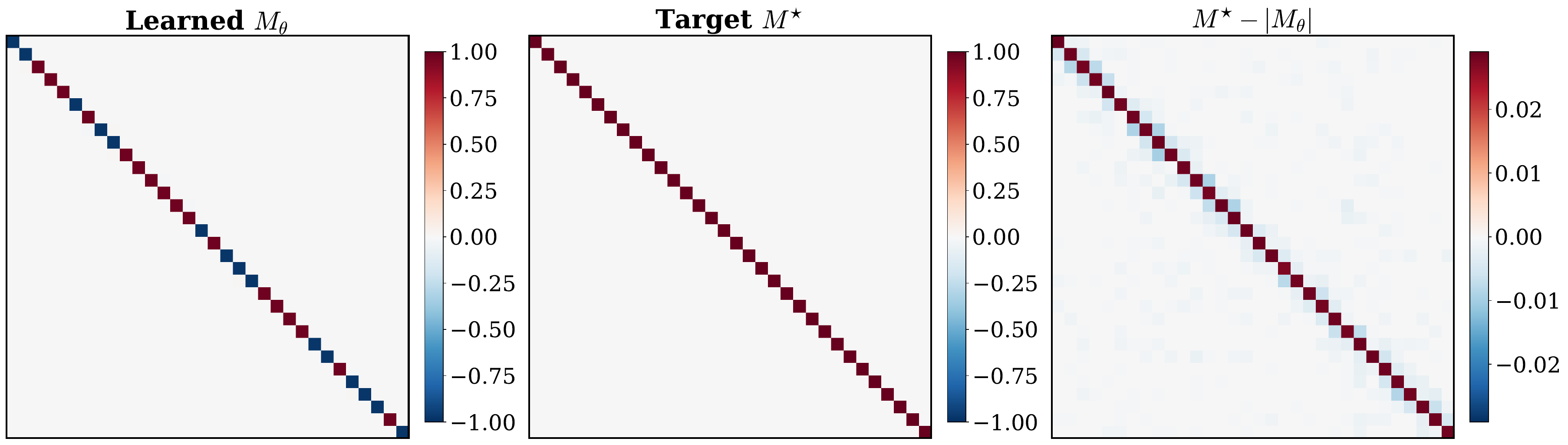}
  \caption{\rev{Learned diffusion matrix $M_\theta$ compared with the analytical target
$M^\star$. Left: learned matrix. Center: analytical discrete diffusion
operator. Right: difference $M^\star - |M_\theta|$. The first $32\times 32$ block
is shown. The learned operator recovers the diagonal structure of the
Identity matrix, although small
residual deviations remain.}}
  \label{fig:M}
\end{figure*}

\rev{In the main text we reported only the drift matrix $K_{\theta}$. For completeness,
figure~\ref{fig:M} shows the noise coupling matrix $M_{\theta} $. The model autonomously recovers the diagonal structure of the noise covariance. Because $M_{\theta}$ multiplies a noise term that is symmetric around zero, $M_{\theta,ii}$ and $-M_{\theta,ii}$ are indistinguishable, and both are valid solutions of the optimization problem.    }

\begin{acknowledgments}
\textit{Acknowledgments}
This project has received funding from the European Union’s Horizon Europe research and innovation programme under the Marie Skłodowska-Curie COFUND grant agreement No 101127936 (DeMythif.AI) and 
by the French ANR grant Scalp (ANR-24-CE23-1320).
\end{acknowledgments}

\bibliographystyle{iopart-num}
\bibliography{paper_refs_publication}

\end{document}